\documentclass[aps,prd,preprint,groupedaddress]{revtex4-1}
\usepackage{amsmath,amssymb}
\usepackage{graphicx}
\begin{document}

\title{A general framework to diagonalize vector--scalar and axial-vector--pseudoscalar transitions in the effective meson Lagrangian}

\author{J. Morais}
\email[]{jorge.m.r.morais@gmail.com}
\thanks{}
\altaffiliation{}
\affiliation{CFisUC, Department of Physics, University of Coimbra, 3004-516 Coimbra, Portugal}

\author{B. Hiller}
\email[]{brigitte@teor.fis.uc.pt}
\altaffiliation{}
\affiliation{CFisUC, Department of Physics, University of Coimbra, 3004-516 Coimbra, Portugal}

\author{A. A. Osipov}
\email[]{osipov@nu.jinr.ru}
\altaffiliation{}
\affiliation{Joint Institute for Nuclear Research, Bogoliubov Laboratory of Theoretical Physics, 141980 Dubna, Russia}

\date{\today}

\begin{abstract}
A new mathematical framework for the diagonalization of the nondiagonal vector--scalar and axial-vector--pseudoscalar mixing in the effective meson Lagrangian is described. This procedure has unexpected connections with the Hadamard product of $n\times n$ matrices describing the couplings, masses, and fields involved. The approach is argued to be much more efficient as compared with the standard methods employed in the literature. The difference is especially noticeable if the chiral and flavor symmetry is broken explicitly. The paper ends with an illustrative application to the chiral model with broken $U(3)_L\times U(3)_R$ symmetry.   
\end{abstract}


\maketitle

\section{Introduction}
The QCD Lagrangian with $n$ massless flavors is known to possess a large global symmetry, namely the symmetry under $U(n)_V\times U(n)_A$ chiral transformations of quark fields. It has been shown by Coleman and Witten \cite{Coleman80} that, in the limit of a large number of colors $N_c$, under reasonable assumptions, this symmetry group must spontaneously break down to the diagonal $U(n)_V$. Consequently, the massless quarks get their constituent masses $M_0$, and massless Goldstone bosons appear in the spectrum \cite{Nambu60,Goldstone61,Goldstone62}. These non-perturbative features of the QCD vacuum can be modeled in analogy with the phenomenon of superconductivity \cite{Nambu61a,Nambu61b}. For that, one should regard the constituent quarks as quasiparticle excitations and the mesons as the bound states of quark-antiquark pairs. The dynamics of such bound states is described by the chiral effective Lagrangian \cite{Kikkawa76,Volkov86,Ebert86,Bijnens93,Bernard96}.

Were the axial $U(n)_A$ symmetry exact, one would observe parity degeneracy of all states with otherwise the same quantum numbers. Due to the mechanism of spontaneous symmetry breaking, described by Nambu and Jona-Lasinio (NJL), the mass splitting occurs between chiral partners, e.g. $m_{a_1}/m_\rho =\sqrt{Z}$, where $Z\simeq 2$ in accordance with a celebrated Weinberg result \cite{Weinberg67,Gilman68,Weinberg69}. In fact, the splitting between $J^{P}=1^{-}$ and $1^{+}$ states is a result of the partial Higgs mechanism: the $A_\mu\partial^\mu\phi$ mixing term which appears in the free meson Lagrangian \cite{Gasiorovicz69} after spontaneous symmetry breaking must be canceled by an appropriate redefinition of the longitudinal component of the massive axial-vector field $A_\mu=A'_\mu +\kappa\partial_\mu\phi$. The result is that the axial field $A_\mu$ ``eats a piece'' of the Goldstone boson $\phi$ and ``gets fat'': $m_{a_1}^2-m_\rho^2=6M^2_0$. Here the quark mass may be expressed in terms of observable values: $f_\pi$ (the pion decay constant) and $g_\rho$ (the $\rho\to\pi\pi$ decay constant), namely $6M_0^2=Zg_\rho^2 f_\pi^2$.   

It is commonly believed that axial-vector fields $A_\mu$, defined in the symmetric vacuum, and $A'_\mu$, defined in the non-symmetric vacuum, should have the same chiral transformations \cite{Ecker89}. As a consequence of that rather natural idea one must use the covariant derivative $\nabla_\mu\phi$ in the replacement above and write $A_\mu=A'_\mu +\kappa\nabla_\mu\phi$. Such derivative contains non-linear field combinations. Thus, upon substitution in the quadratic form to be diagonalized, new meson interaction terms emerge which are not present in the original Lagrangian. Although, in principle, there is no reason to object to new interactions as long as they fulfill the symmetry requirements, they are subleading in large $N_c$ counting as compared to the mixing which occurs at leading order. If one restricts the analysis to leading order in $N_c$, i.e. to the level of the free meson Lagrangian, chiral symmetry will not be supported.

Recently \cite{Volkov17,Osipov17}, it has been shown that the linear replacement $A_\mu=A'_\mu +\kappa\partial_\mu\phi$ changes the chiral transformation laws of the axial-vector field $A'_\mu$ as compared to the $A_\mu$ ones, though in a way that leaves the group properties intact. The special thing about these new transformations is their dependence on the classical parameter $M'$ which is a diagonal $n\times n$ matrix in the $n$-flavor space; in the non-symmetric ground state its eigenvalues are all equal and nonzero $M'=\mbox{diag}(M_0,M_0,\ldots ,M_0)$; in the symmetric vacuum, $M_0=0$ and the transformations coincide with the standard ones.  

In nature, chiral symmetry is also broken explicitly by the current quark masses $m =\mbox{diag}(m_u, m_d, m_s)$ (it has been shown in \cite{Vafa84} for QCD with two flavors that isospin is not spontaneously broken; in the following, we will consider mainly the case $n=3$, although our discussion is valid for an arbitrary number of flavors $n$). Due to the current quark masses, the $U(n)_V$ symmetry breaks down to $U(1)_V^n$. Then, it follows from the gap equation that the constituent quark mass matrix $M$ is diagonal but its eigenvalues are all unequal and nonzero $M=\mbox{diag}(M_u,M_d,M_s)$. This leads to a new mixing between the vector, $V_\mu$, and the scalar, $\sigma$, fields and, as a result, to the redefinition of the longitudinal component of the vector field: $V_\mu=V'_\mu +\kappa'\partial_\mu\sigma$. This makes the vector field $V'_\mu$ heavier. 

It is the purpose of this letter to point out that the result \cite{Volkov17,Osipov17} can be extended to the realistic case of $M=\mbox{diag}(M_u,M_d,M_s)$. To be precise, we obtain a new sort of infinitesimal chiral transformations of spin-0 and spin-1 fields in the non-symmetric ground state when flavor symmetry is broken explicitly (this is the main result of our work). Surprisingly, the effects of flavor symmetry breaking, collected in the matrix $M$, do not spoil the $U(3)\times U(3)$ group transformation laws of the fields, although $M$ enters the transformations. It shows that a linear replacement of variables that diagonalizes the free part of the meson Lagrangian is legitimate, unique, and does not ruin the pattern of explicit symmetry breaking of the theory. Indeed, (a) it is legitimate because this replacement does not lead to chiral symmetry breaking in the Lagrangian, although it changes the chiral transformation properties of the spin-1 fields in the non-symmetric ground state. (b) It is unique from the point of view of the $1/N_c$ expansion because it solves the problem of diagonalization already at leading-$N_c$ order, i.e. at the level of the free Lagrangian. (It is well-known that mesons for large $N_c$ are free, stable, and non-interacting. Meson decay amplitudes are of order $1/\sqrt{N_c}$, and meson-meson elastic scattering amplitudes are of order $1/N_c$ \cite{Hooft74a,Hooft74b,Witten79}.) All other approaches \cite{Gasiorovicz69,Meissner88,Ecker89,Bijnens93,Birse96,Osipov00} give the same result at this order. (c) The replacement of variables obtained in this work preserves the pattern of chiral symmetry breaking because the transformations belong to the $U(3)\times U(3)$ group. Thus, the replacement does not generate a new contribution to the divergence of the axial-vector current. 
    
We also demonstrate that the linear replacement of variables that we found has unexpected connections with the Hadamard product of $n\times n$ matrices describing the couplings, masses, and fields. That allows us to reformulate the diagonalization procedure entirely in terms of the Hadamard product, in contrast to the conventional methods used in the literature which we refer to as being standard.  

To find the chiral transformations of the meson fields in the non-symmetric vacuum we use the NJL Lagrangian which includes both spin-0 and spin-1 $U(3)\times U(3)$ symmetric four-quark interactions. It is known that this Lagrangian undergoes dynamical symmetry breaking \cite{Hatsuda94}. It also reproduces the qualitative features of the large-$N_c$ limit. The model describes both phases and gives a solid framework for the study of the transformation laws of the $q\bar q$ bound states - mesons. Indeed, after some standard redefinitions, one can track the chiral transformation properties of the fields starting from fundamental quarks in the Wigner--Weyl phase and ending up with quark-antiquark bound states in the Nambu--Goldstone phase. 

The source of interest regarding the chiral transformations of spin-1 fields resides in the use of these fields in the effective Lagrangians describing the strong interactions of hadrons at energies of $\sim 1\,\mbox{GeV}$ \cite{Meissner88,Bando88,Birse96}. Presently these theories are actively used in studies of $\tau$-lepton decay modes \cite{Volkov2012tau}, $e^+ e^-$ hadron production \cite{Volkov2012elpos}, and the QCD phase diagram \cite{Benic2014,Chu2015}. The chiral transformations we suggest allow for greater freedom in the construction of such Lagrangians by noting that one may use different representations for these transformations as long as they obey the same group structure. In this context it is important that transformations include the flavor symmetry breaking effects while not spoiling the symmetry breaking pattern of the theory. The latter is very important for controllable calculations.

In Section 2 we study the NJL model of quarks with the $U(3)\times U(3)$ symmetric four-quark interactions and obtain the linear chiral transformations of the spin-0 and spin-1 meson fields in the symmetric ground state. In Section 3 we deal with the effective meson Lagrangian in the non-symmetric phase. Here we discuss the $A-\phi$ and $V-\sigma$ diagonalization and outline the construction of new chiral transformation laws for spin-1 fields. This section contains our main result. In Section 4, as an application, we consider an extended model with four- and eight-quark interactions which include the explicit symmetry breaking vertices. This sophisticated example clearly shows the efficiency of the approach based on the Hadamard product method. On the other hand, the model may have some interest for the readers who might wish to use it to take into account the explicit and flavor symmetry breaking effects in hot/dense/magnetized matter \cite{Gatto2010,Moreira2014}, or apply it in the study of hybrid stars \cite{Pais2016}, where eight-quark interactions seem to have an important impact.  

\section{Chiral Transformations of Meson Fields}
We start from the quark version of the original NJL model \cite{Nambu61a,Nambu61b} with nonlinear four-quark spin-0 and spin-1 interactions which allow a chiral group $G=U(3)_V\times U(3)_A$ \cite{Volkov86,Ebert86,Bijnens93}, and where there is considerable freedom in the choice of auxiliary fields in the vector and axial-vector channels. The Lagrangian density of the model is 
\begin{equation}
\label{lag}
{\cal L}=\bar q(i\gamma^\mu\partial_\mu - m)q + \frac{G_S}{2}\left[(\bar q\lambda_aq)^2+(\bar qi\gamma_5\lambda_a q)^2 \right] - \frac{G_V}{2}\left[(\bar q\gamma^\mu\lambda_a q)^2+(\bar q\gamma^\mu\gamma_5\lambda_a q)^2 \right]. 
\end{equation}
$G_S$ and $G_V$ are universal four-quark coupling constants with dimensions (length)$^2$ and $m$ is the current quark mass matrix. The symmetry group $G$ acts on the quark fields $q$ (in this notation the color and flavor indices are suppressed) as follows
\begin{equation}
q \rightarrow q' = e^{i \left(\alpha + \gamma_5 \beta\right)} q, \qquad \bar{q} \rightarrow \bar{q}' = \bar{q} e^{-i \left(\alpha - \gamma_5 \beta\right)}.
\end{equation}
Here $\alpha = \alpha_a\lambda_a/2$, $\beta = \beta_a\lambda_a/2$, $a=0,1,\ldots,8$ and $\alpha_a, \beta_a \in \mathbb{R}$; $\lambda_0=\sqrt{\frac23}\,\bf 1$ with $\bf 1$ being a unit $3\times 3$ matrix and $\lambda_a$ $(a>0)$ are the usual $SU\left(3\right)$ Gell-Mann matrices with the following basic trace property $\mbox{tr}(\lambda_a\lambda_b)=2\delta_{ab}$; the resulting infinitesimal transformations are
\begin{equation}
\label{quark-transfs}
\delta q = q' - q = i \left(\alpha + \gamma_5 \beta\right) q, \qquad \delta \bar{q} = \bar{q}' - \bar{q} = -i \bar{q}\left(\alpha - \gamma_5 \beta\right).
\end{equation}
This symmetry is explicitly broken due to non-zero values of the current quark masses $m=\mbox{diag}(m_u,m_d,m_s)$, i.e. we have 
\begin{equation}
\label{esb}
\delta {\cal L}=-\delta\left(\bar qmq\right)=i\bar q\left([\alpha, m]-\gamma_5\{\beta, m\}\right)q.
\end{equation}

Following \cite{Kikkawa76}, one may wish to introduce auxiliary fields $\sigma_a =G_S(\bar q\lambda_aq)$, $\phi_a=G_S(\bar qi\gamma_5\lambda_aq)$, $V_{\mu a}=G_V(\bar q\gamma_\mu\lambda_aq)$, $A_{\mu a}=G_V(\bar q\gamma^\mu\gamma_5\lambda_a q)$, and resort equivalently (in the functional integral sense) to the theory with the Lagrangian density 
\begin{equation}
\label{quark-det}
{\cal L'}=\bar{q} D q -\frac{1}{4G_S}\mbox{tr}\left[(\sigma +m)^2+\phi^2\right]+\frac{1}{4G_V}\mbox{tr}\left(V_\mu^2+A_\mu^2\right),
\end{equation}
with $D$ being a Dirac operator in the presence of mesonic fields 
\begin{equation}
D = i\gamma^{\mu}\partial_{\mu} - \sigma - i\gamma_5 \phi - \gamma^{\mu} V_{\mu} - \gamma^{\mu} \gamma_5 A_{\mu} ,
\end{equation}
$\sigma =\sigma_a\lambda_a$, $\phi =\phi_a\lambda_a$, $V_\mu =V_{\mu a}\lambda_a$, $A_\mu =A_{\mu a}\lambda_a$, and where the trace is to be taken in flavor space. 

Perhaps one should explain here how $m$ arises in the meson part of the Lagrangian density (\ref{quark-det}). It is a trivial result following from the standard replacements of variables in the corresponding functional integral. Nonetheless, one may doubt if the explicit symmetry breaking pattern does not change. From (\ref{quark-det}) one can see that it does not  
\begin{equation}
\label{sbp}
\delta {\cal L'}=-\frac{1}{2G_S}\mbox{tr}\left(m\delta\sigma\right)=-\frac{\delta\sigma_a}{2G_S}\mbox{tr}(m\lambda_a)=-\frac{1}{G_S}m_a\delta\sigma_a=-\delta\left(\bar qmq\right)=\delta {\cal L},    
\end{equation}
where $m_a$ is defined by $m_a=\frac{1}{2}\mbox{tr}(m\lambda_a)$ with $m=m_a\lambda_a =\mbox{diag}(m_u,m_d,m_s)$.

The latter implies that
\begin{equation}
\label{chi-inv-cond}
\delta \left(\bar{q} D q\right) = 0,
\end{equation}
or equivalently
\begin{equation}
\label{chi-inv-cond-2}
\delta \left\lbrack\bar{q} \left( \sigma + i\gamma_5 \phi \right) q\right\rbrack = 0, \qquad \delta \left\lbrack\bar{q} \left( \gamma^{\mu} V_{\mu} + \gamma^{\mu} \gamma_5 A_{\mu} \right) q\right\rbrack = 0.
\end{equation}
The symmetry is now
\begin{align}
\label{sigma-transf}
\delta \sigma & = i \left\lbrack \alpha, \sigma \right\rbrack + \left\lbrace \beta, \phi \right\rbrace , \\
\label{phi-transf}
\delta \phi & = i \left\lbrack \alpha, \phi \right\rbrack - \left\lbrace \beta, \sigma \right\rbrace ,\\
\label{V-transf}
\delta V_{\mu} & = i \left\lbrack \alpha, V_{\mu} \right\rbrack + i \left\lbrack \beta, A_{\mu} \right\rbrack ,\\
\label{A-transf}
\delta A_{\mu} & = i \left\lbrack \alpha, A_{\mu} \right\rbrack + i \left\lbrack \beta, V_{\mu} \right\rbrack.
\end{align}

These transformation laws are valid in the symmetric Wigner-Weyl realization of chiral symmetry, where $\left\langle \sigma \right\rangle = 0$. The generators possess a Lie algebra structure
\begin{eqnarray}
\label{J}
&&\delta_{[1,2]}q=i\left(\alpha_{\left\lbrack 1,2 \right\rbrack}+\gamma_5\beta_{\left\lbrack 1,2 \right\rbrack}\right) q=[\delta_1, \delta_2]q, \nonumber \\
&&\delta_{[1,2]}\sigma = i\left\lbrack \alpha_{\left\lbrack 1,2 \right\rbrack},\sigma\right\rbrack + \left\lbrace \beta_{\left\lbrack 1,2 \right\rbrack},\phi\right\rbrace = [\delta_1, \delta_2]\sigma, \nonumber \\
&&\delta_{[1,2]}\phi = i\left\lbrack \alpha_{\left\lbrack 1,2 \right\rbrack},\phi\right\rbrack - \left\lbrace \beta_{\left\lbrack 1,2 \right\rbrack},\sigma\right\rbrace = [\delta_1, \delta_2]\phi, \nonumber \\
&&\delta_{[1,2]}V_\mu = i\left\lbrack \alpha_{\left\lbrack 1,2 \right\rbrack},V_\mu\right\rbrack + i\left\lbrack \beta_{\left\lbrack 1,2 \right\rbrack},A_\mu\right\rbrack = [\delta_1, \delta_2]V_\mu , \nonumber \\
&&\delta_{[1,2]}A_\mu = i\left\lbrack \alpha_{\left\lbrack 1,2 \right\rbrack},A_\mu\right\rbrack + i\left\lbrack \beta_{\left\lbrack 1,2 \right\rbrack},V_\mu\right\rbrack = [\delta_1, \delta_2]A_\mu , 
\end{eqnarray} 
where 
\begin{equation}
\label{alpha12}
i\alpha_{\left\lbrack 1,2 \right\rbrack}=\left\lbrack \alpha_1,\alpha_2\right\rbrack + \left\lbrack \beta_1,\beta_2\right\rbrack, \qquad i\beta_{\left\lbrack 1,2 \right\rbrack} =\left\lbrack \alpha_1,\beta_2\right\rbrack + \left\lbrack \beta_1,\alpha_2\right\rbrack.
\end{equation} 

\section{$V\!-\!\sigma$ and $A\!-\!\phi$ Mixings and General Linear Shift}
\label{vs}
For further progress we have to bosonize the theory to get profit from the $1/N_c$ expansion. In the large-$N_c$ limit the meson functional integral of the theory (\ref{quark-det}) is dominated by the stationary phase (or mean-field) configurations $\sigma =M, \phi =V_\mu=A_\mu=0$. The elements of the diagonal matrix $M$ differ from zero and may be interpreted as constituent quark masses $M=\mbox{diag}\,(M_u,M_d,M_s)$. The latter is a consequence of the gap equations
\begin{equation}
\label{gap}
M_i-m_i=i8G_SN_c\int_\Lambda\frac{d^4k}{(2\pi )^4}\frac{M_i}{k^2-M_i^2}=\frac{N_cG_S}{2\pi^2}M_i\left[\Lambda^2-M_i^2\ln\left(1+\frac{\Lambda^2}{M_i^2}\right)\right],   
\end{equation}  
where $i=u,d,s$ and $\Lambda$ is an intrinsic cutoff of the NJL model. The current quark masses $m_i$ affect (through the gap equation) the constituent quark masses $M_i$ which accumulate the explicit and flavor symmetry breaking effects enhancing them. In particular, in the chiral limit $m_i=0$, it follows that either $M_i\to M_0=0$ (if $G_SN_c\Lambda^2/(2\pi^2)<1$, Wigner-Weyl phase) or $M_i\to M_0>0$ (if $G_SN_c\Lambda^2/(2\pi^2)>1$, Nambu-Goldstone phase). 

Since a dynamically broken symmetry is not spoiled in the Lagrangian, we can expand around a non-zero vacuum expectation value of $\sigma$ without breaking the symmetry. For that we perform the shift $\sigma \rightarrow \sigma + M$, leading to 
\begin{equation}
D\to D_M = i\gamma^{\mu}\partial_{\mu} - \left(\sigma + M\right) - i\gamma_5 \phi - \gamma^{\mu} V_{\mu} - \gamma^{\mu} \gamma_5 A_{\mu}.
\end{equation}

The equation (\ref{chi-inv-cond}) must still hold, now in the form 
\begin{equation}
\label{detM}
\delta (\bar qD_Mq)=0 ,
\end{equation}
because otherwise the symmetry breaking pattern (\ref{sbp}) will be not preserved. This gives us the modified transformation laws of spin-0 fields in the non-symmetric phase
\begin{align}
\label{sigma-transf-2}
\delta \sigma & = i \left\lbrack \alpha, \sigma + M \right\rbrack + \left\lbrace \beta, \phi \right\rbrace , \\
\label{phi-transf-2}
\delta \phi & = i \left\lbrack \alpha, \phi \right\rbrack - \left\lbrace \beta, \sigma + M \right\rbrace.
\end{align}

We can still associate the infinitesimal transformations (\ref{sigma-transf-2},\ref{phi-transf-2}) with the chiral group $G$, because $M$ does not ruin the symmetry algebra of $G$ given by (\ref{J},\ref{alpha12}). This can be easily checked. On the other hand, the existence of such a symmetry shows that the part of the effective meson Lagrangian following from $\bar qD_Mq$ (after integrating over the quark degrees of freedom in the generating functional of the theory) includes vertices with explicit and flavor symmetry breaking effects which are still altogether G-invariant as expressed in (\ref{detM}).

The results above have no consequences for the transformations of spin-1 fields in the non-symmetric phase. The equations (\ref{V-transf}) and (\ref{A-transf}) agree with the requirement (\ref{detM}). 

Let us now turn to the vacuum-to-vacuum transition amplitude of the model beyond the mean-field approximation
\begin{eqnarray}
S&=&\int\!{\cal D}\sigma_a {\cal D}\phi_a{\cal D}V^\mu_a{\cal D}A^\mu_a\exp i\!\int\! d^4x \left\{\frac{1}{4G_V}\mbox{tr}\left(V_\mu^2+A_\mu^2\right)
 - \frac{1}{4G_S}\mbox{tr}\left[(\sigma +M+m)^2+\phi^2\right]\right\} \nonumber \\
&\times&\int\! {\cal D}q{\cal D}\bar q \exp i\!\int\! d^4x \left( \bar q D_Mq
\right). 
\end{eqnarray} 
To obtain the effective meson Lagrangian one should consider the long wavelength expansion of the quark determinant, $\mbox{det}D_M$, the formal expression for the path integral over quarks. The appropriate tool here is the Schwinger-DeWitt method \cite{Ball89}. This yields the local low-energy effective meson Lagrangian. Renormalizing the meson fields by bringing their kinetic terms to the standard form (e.g., $\phi^R=g\phi$, where $g\sim 1/\sqrt{N_c}$), one arrives at the picture corresponding to the large $N_c$ limit: the free parts of the meson Lagrangian count as $g^2N_c\sim N_c^0$, the three-meson interactions as $g^3N_c\sim 1/\sqrt{N_c}$, the four-meson amplitudes as $g^4N_c\sim 1/N_c$ \cite{Volkov86,Ebert86}. 

Obviously, the $V\!-\!\sigma$ and $A\!-\!\phi$ mixing arises at $N_c^0$ order: the mixing is described by vertices proportional to $\mbox{tr}\left(A_\mu\{M,\partial^\mu\phi\}\right)$ and $\mbox{tr}\left(V_\mu [M,\partial^\mu\sigma ]\right)$. The first one is a result of spontaneous symmetry breaking, while the second one is a direct consequence of the flavor symmetry breaking enforced in the broken phase. Both lead to additional contributions to the kinetic terms of pseudoscalar (through the transitions $\partial\phi\to A\to\partial\phi$) and scalar (through the transitions $\partial\sigma\to V\to\partial\sigma$) states. Consequently, these fields must be renormalized again to the standard form. In this case one gets correct expressions for the masses of spin-0 states. 

Alternatively, one may wish to eliminate the mixing and diagonalize the free Lagrangian by the replacements
\begin{equation}
\label{shifts}
V_{\mu}=V_{\mu}' + X_{\mu}, \qquad A_{\mu}=A_{\mu}' + Y_{\mu},
\end{equation}
where the entries of $X_{\mu}$ and $Y_{\mu}$ are appropriate combinations of spin-0 fields. They should also depend on $M$, as it is required by the mixing terms. These replacements introduce further mixing terms through the $V_{\mu}$ and $A_{\mu}$ mass terms which must add up to zero in the end, fixing the coefficients of the combinations used in (\ref{shifts}). In principle, there are an infinity of possible physically equivalent replacements in the form of (possibly infinite) sums of field products. According to Chisholm's theorem \cite{Chisholm61,Salam61}, all such redefinitions yield the same result when computing observables as long as they preserve the form of the free part of the Lagrangian. This reasoning ensures us that we may always restrict to the minimal necessary terms for our intended purposes, i.e. to linear field combinations. 

From the point of view of the $1/N_c$ expansion, the minimal replacement in (\ref{shifts}) is unique. All others include additional nonlinear combinations in fields, but must coincide with (\ref{shifts}) in their linear part, i.e. at $N_c^0$ order (see also discussion around eq.26). This is a direct consequence of the fact that the mixing terms have their origin at the level of the free Lagrangian.

We may now require that the replacement (\ref{shifts}) does not violate the symmetry condition (\ref{detM}). This is ensured if the spin-1 states transform like 
\begin{equation}
\label{vac}
\delta \left\lbrace \bar{q} \left\lbrack \gamma^{\mu} \left(V_{\mu}' + X_{\mu}\right) + \gamma^{\mu} \gamma_5 \left(A_{\mu}' + Y_{\mu}\right)\right\rbrack q \right\rbrace =0
\end{equation}
Gathering separately the factors multiplying $\gamma^{\mu}$ and $\gamma^{\mu} \gamma_5$ we conclude that (\ref{vac}) is equivalent to
\begin{align}
\label{V-shift-transf}
\delta V_{\mu}' & = i \left\lbrack \alpha, V_{\mu}' + X_{\mu} \right\rbrack + i \left\lbrack \beta, A_{\mu}' + Y_{\mu} \right\rbrack - \delta X_{\mu}, \\
\label{A-shift-transf}
\delta A_{\mu}' & = i \left\lbrack \alpha, A_{\mu}' + Y_{\mu} \right\rbrack + i \left\lbrack \beta, V_{\mu}' + X_{\mu} \right\rbrack - \delta Y_{\mu}.
\end{align}
These transformations must preserve the algebraic structure of the chiral group $G$, i.e. the composition properties of the group which are specified in (\ref{alpha12}). 

It seems natural to require that $V_\mu'$ and $A_\mu'$ have the same transformation properties as $V_\mu$ and $A_\mu$. From (\ref{V-shift-transf})-(\ref{A-shift-transf}), it follows then that $X_\mu$ and $Y_\mu$ need to be chiral partners and should also transform like $V_\mu$ and $A_\mu$. This would disable the linear solution $X_\mu\propto\partial_\mu\sigma$ and $Y_\mu\propto\partial_\mu\phi$ due to the transformation properties of spin-0 fields (\ref{sigma-transf-2}) and (\ref{phi-transf-2}). Instead, one would look for nonlinear combinations of fields for $X_\mu$ and $Y_\mu$ which respect the symmetry transformations (\ref{V-transf}) and (\ref{A-transf}) and contain the linear terms necessary for the diagonalization. For instance, if we assume that flavor symmetry is unbroken, we may use the solution \cite{Osipov00},
\begin{equation}
\label{nonlin}
X_\mu = -i\kappa\left( \left[\sigma +M,\partial_\mu\sigma\right]+\left[\phi ,\partial_\mu\phi\right]\right), \qquad Y_\mu=\kappa\left(\left\{\sigma +M,\partial_\mu\phi\right\}-\left\{\phi , \partial_\mu\sigma\right\}\right),
\end{equation}
where the constant $\kappa$ is fixed by a diagonalization condition. The nonlinear terms essentially modify the interaction part of an effective meson Lagrangian without physical consequences \cite{Chisholm61,Volkov17}. However, if the flavor symmetry is broken the replacement (\ref{nonlin}) does not solve the problem. This is why such natural replacements are not efficient in the direct calculations.    

Now, let us consider the minimal replacement in (\ref{shifts}). In this case, we are faced with the opposite situation: it makes calculations as simple as possible, but one should take care in justifying such a replacement. Indeed, in this case, as it follows from (\ref{V-shift-transf})-(\ref{A-shift-transf}), the fields $V_\mu'$ and $A_\mu'$ cannot transform like $V_\mu$ and $A_\mu$. Can this introduce some spurious symmetry breaking and change the physics of spin-1 states? To answer the question we compute the commutators $\delta_{[1,2]}V_\mu'$ and $\delta_{[1,2]}A_\mu'$. The symmetry will be respected if the commutators depend only on the parameters of the infinitesimal chiral transformations $\alpha_{[1,2]}$ and $\beta_{[1,2]}$, i.e., leaving the group composition properties (\ref{alpha12}) unchanged. We have  
\begin{eqnarray}
&[\delta_1, \delta_2]V_{\mu}'=i \left\lbrack \alpha_{\left\lbrack 1,2 \right\rbrack}, V_{\mu}' + X_{\mu} \right\rbrack + i \left\lbrack \beta_{\left\lbrack 1,2 \right\rbrack}, A_{\mu}' + Y_{\mu} \right\rbrack - [\delta_1, \delta_2] X_{\mu}, \\
&[\delta_1, \delta_2]A_{\mu}'=i \left\lbrack \alpha_{\left\lbrack 1,2 \right\rbrack}, A_{\mu}' + Y_{\mu} \right\rbrack + i \left\lbrack \beta_{\left\lbrack 1,2 \right\rbrack}, V_{\mu}' + X_{\mu} \right\rbrack - [\delta_1, \delta_2] Y_{\mu}.
\end{eqnarray}
One can see that the chiral group structure will be preserved overall as long as $X_{\mu}$ and $Y_{\mu}$ can be chosen in such a way that their transformation laws obey the algebraic structure of $G$ as well
\begin{equation}
\label{asc}
[\delta_1,\delta_2] X_{\mu}=\delta_{\left\lbrack 1,2 \right\rbrack} X_{\mu}, \qquad [\delta_1,\delta_2] Y_{\mu}=\delta_{\left\lbrack 1,2 \right\rbrack} Y_{\mu}.
\end{equation}
This clearly defines the freedom one may have in the choices of $X_\mu$ and $Y_\mu$. In particular, it is not forbidden for them to transform like spin-0 chiral partners $X_\mu\sim\partial_\mu\sigma$ and $Y_\mu\sim\partial_\mu\phi$ as it follows from the diagonalization procedure of the considered NJL model. Another interesting case has been considered in \cite{Volkov17,Osipov17}, where $X_\mu =0$, but $Y_\mu\neq 0$ (the case without flavor symmetry breaking).      
 
Any point made so far on the chiral transformation properties of fields in matrix form may be carried over to a formulation based on individual matrix entries. If the Lagrangian contains mixing terms in the form $\text{tr}\left(V^{\mu} \left\lbrack M,\partial_{\mu}\sigma \right\rbrack\right) = \text{tr}\left(V^{\mu}\left(i\Delta_M \circ \partial_{\mu} \sigma\right)\right)$ and $\text{tr}\left(A^{\mu} \left\lbrace M,\partial_{\mu}\phi \right\rbrace\right) = \text{tr}\left(A^{\mu}\left(\Sigma_M \circ \partial_{\mu} \phi\right)\right)$, it suffices to define $X_\mu$ and $Y_\mu$ as
\begin{equation}
\label{shifts-2}
X_\mu = k\circ\Delta_M\circ\partial_\mu\sigma, \qquad Y_\mu=k'\circ\Sigma_M\circ \partial_\mu\phi .
\end{equation}
Here, $k$ and $k'$ are symmetric coefficient matrices in a flavor space whose entries should be fixed from the Lagrangian diagonalization requirements; $\Delta_M$ and $\Sigma_M$ are mass-dependent matrices defined as
\begin{equation}
\label{Delta-Sigma-defs}
\left(\Delta_M\right)_{ij} = -i \left(M_i - M_j\right), \qquad \left(\Sigma_M\right)_{ij} = M_i + M_j,
\end{equation}
and the symbol $\circ$ stands for the Hadamard (or Schur) product (see e.g. \cite{Styan73}) defined as
\begin{equation}
\left(A \circ B\right)_{ab} = A_{ab} B_{ab},
\end{equation}
without summation over repeated indices. This product is commutative unlike regular matrix multiplication, but the associative property is retained, as well as the distributive property over matrix addition, i.e.
\begin{eqnarray}
& A \circ B =  B \circ A, \nonumber \\
& A \circ B \circ C = \left(A \circ B\right) \circ C = A \circ \left(B \circ C\right), \nonumber \\
& A \circ \left(B + C\right) = A \circ B + A \circ C.
\end{eqnarray}

Definitions (\ref{shifts-2}) yield the following transformation laws for the shifted longitudinal components of $V_{\mu}$ and $A_{\mu}$ fields
\begin{eqnarray}
\delta X_{\mu}&=& k\circ\Delta_M\circ\delta\partial_\mu\sigma = k\circ\Delta_M\circ \left( i \left\lbrack \alpha, \partial_{\mu} \sigma \right\rbrack + \left\lbrace \beta, \partial_{\mu} \phi \right\rbrace \right), \\
\delta Y_{\mu}&=& k'\circ\Sigma_M\circ\delta\partial_\mu\phi = k'\circ\Sigma_M\circ \left( i \left\lbrack \alpha, \partial_{\mu} \phi \right\rbrack - \left\lbrace \beta, \partial_{\mu} \sigma \right\rbrace \right),
\end{eqnarray}
and their Lie brackets yield
\begin{eqnarray}
&&[\delta_1,\delta_2]X_\mu = k\circ\Delta_M\circ\left( i\left\lbrack\alpha_{\left\lbrack 1,2 \right\rbrack},\partial_{\mu}\sigma\right\rbrack + \left\lbrace\beta_{\left\lbrack 1,2 \right\rbrack},\partial_\mu\phi\right\rbrace\right) =
\delta_{\left\lbrack 1,2 \right\rbrack} X_{\mu}, \\
&&[\delta_1,\delta_2]Y_\mu= k'\circ\Sigma_M\circ\left( i \left\lbrack \alpha_{\left\lbrack 1,2 \right\rbrack}, \partial_{\mu} \phi \right\rbrack - \left\lbrace \beta_{\left\lbrack 1,2 \right\rbrack}, \partial_{\mu} \sigma \right\rbrace \right)=
\delta_{\left\lbrack 1,2 \right\rbrack} Y_\mu, 
\end{eqnarray}
with $\alpha_{\left\lbrack 1,2 \right\rbrack}$, $\beta_{\left\lbrack 1,2 \right\rbrack}$ defined in (\ref{alpha12}). We see that conditions (\ref{asc}) are fulfilled. Thus, eqs. (\ref{V-shift-transf},\ref{A-shift-transf}) are the hidden symmetry transformations of $G$ in the broken vacuum. It means that the minimal replacement of variables has no physical consequences as compared to the standard nonlinear replacement even if the flavor symmetry is broken.   

\section{Application to an $SU\left(3\right)$ Chiral Model}
Now, let us consider a physical application of the method presented above. For that we chose a recently proposed effective $U(3)\times U(3)$ chiral model \cite{Morais17}. It extends the model \cite{Ebert86} presented in the text by including the $U\left(1\right)_A$ breaking 't Hooft interaction \cite{tHooft76}, eight-quark interactions and systematically taking into account the explicit and flavor symmetry breaking effects. Our choice is motivated by the growing interest in the eight-quark interactions in hadronic matter, including the physics of stars, and by the importance of the axial anomaly and the flavor symmetry breaking effects in the study of the QCD phase diagram.    

In both models \cite{Ebert86,Morais17} we arrive, after bosonization, to the same mixing terms
\begin{equation}
\label{mixl}
{\cal L}_{mix}=\frac{1}{2\varrho^2} \text{tr} \left(-i \left\lbrack M, V^{\mu} \right\rbrack \partial_{\mu} \sigma -\left\lbrace M, A^{\mu} \right\rbrace \partial_{\mu} \phi \right).
\end{equation}
Here, the trace is to be taken in flavor space; $\varrho^2$ is a constant cutoff dependent factor related with the evaluation of the quark determinant, and $M$ is the constituent quark mass matrix, $M=\mbox{diag}(M_u,M_d,M_s)$, related through the reduced Schwinger-Dyson equation (\ref{gap}) with the current quark masses $m$. 

Eq.(\ref{mixl}) can be recast into a somewhat more explicit form if we make use of the simple relations $[M,A]=i\Delta_M\circ A$ and $\{M,A\}=\Sigma_M\circ A$ which are fulfilled for the diagonal matrix $M$ and any matrix $A=A_a\lambda_a$, where $\lambda_a$ are the hermitian generators of the flavor $U(3)$ group. This gives   
\begin{eqnarray}
{\cal L}_{mix}&=&\frac{1}{2\varrho^2} \text{tr} \left[ \left(\Delta_M \circ V^{\mu}\right) \partial_{\mu} \sigma - \left(\Sigma_M \circ A^{\mu}\right) \partial_{\mu} \phi \right] \nonumber \\
&=& -\frac{1}{2\varrho^2} \text{tr} \left[ \left(\Delta_M \circ \partial_{\mu} \sigma\right) V^{\mu} + \left(\Sigma_M \circ \partial_{\mu} \phi\right) A^{\mu} \right],
\end{eqnarray}
with $\Delta_M,\Sigma_M$ as defined in (\ref{Delta-Sigma-defs}). For simplifying this expression we have used the fact that, for any $U(3)$ symmetric matrix $S$ (e.g. $\Sigma_M$) or antisymmetric matrix $\Omega$ (e.g. $\Delta_M$) and any other $B,C\in U(3)$, it is always true that
\begin{eqnarray}
&&\text{tr}\left[B\left(S\circ C\right)\right]=\sum_{i,j}B_{ij}S_{ji}C_{ji}=\sum_{i,j}B_{ij}S_{ij}C_{ji}=\text{tr}\left[(B\circ S)C\right], \\
&&\text{tr}\left[B\left(\Omega\circ C\right)\right]=\sum_{i,j}B_{ij}\Omega_{ji}C_{ji}=-\sum_{i,j}B_{ij}\Omega_{ij}C_{ji}=-\text{tr}\left[(B\circ \Omega)C\right].
\end{eqnarray}

This form for the mixing terms provides a direct hint to the adequacy of the forms (\ref{shifts-2}) for diagonalizing the Lagrangian. Carrying on such replacements will induce from $V^{\mu}$ and $A^{\mu}$ new similarly shaped mixing terms with $k,k'$ appearing as adjustable coefficients.

The model's mass terms in the unshifted Lagrangian may be expressed as 
\begin{eqnarray}
\label{V-mass}
{\cal L}_V&=&\frac{1}{2\varrho^2} \text{tr} \left(\varrho^2 V^{\mu} \left(H^{\left(1\right)} \circ V_{\mu}\right) - \frac{1}{2} \left\lbrack M, V^{\mu} \right\rbrack \left\lbrack M, V_{\mu} \right\rbrack \right) \nonumber \\
&=&\frac{1}{2\varrho^2} \text{tr} \left(\varrho^2 V^{\mu} \left(H^{\left(1\right)} \circ V_{\mu}\right) + \frac{1}{2} \left(\Delta_M \circ V^{\mu}\right) \left(\Delta_M \circ V_{\mu}\right) \right)
\end{eqnarray}
for $V^{\mu}$ and
\begin{eqnarray}
\label{A-mass}
{\cal L}_A&=&\frac{1}{2\varrho^2} \text{tr} \left(\varrho^2 A^{\mu} \left(H^{\left(2\right)} \circ A_{\mu}\right) + \frac{1}{2} \left\lbrace M, A^{\mu} \right\rbrace \left\lbrace M, A_{\mu} \right\rbrace \right) \nonumber \\
&=& \frac{1}{2\varrho^2} \text{tr} \left(\varrho^2 A^{\mu} \left(H^{\left(2\right)} \circ A_{\mu}\right) + \frac{1}{2} \left(\Sigma_M \circ A^{\mu}\right) \left(\Sigma_M \circ A_{\mu}\right) \right)
\end{eqnarray}
for $A^{\mu}$. Here, $H^{\left(1\right)}$ and $H^{\left(2\right)}$ are symmetric $U(3)$ matrices. In the model \cite{Ebert86} $H^{\left(1\right)}=H^{\left(2\right)}\propto \bf 1$. The model \cite{Morais17} leads to a more general form of $H^{\left(1\right)}$ and $H^{\left(2\right)}$ which makes the standard diagonalization procedure algebraically heavier.

After replacing the fields according to (\ref{shifts-2}), we get the following additional mixing terms:
\begin{equation}
\Delta {\cal L}_{mix}^{V'\sigma}=\frac{1}{2\varrho^2} \text{tr} \left[2\varrho^2 \left(H^{\left(1\right)} \circ k \circ \Delta_M \circ \partial_{\mu} \sigma\right) V^{'\mu} - \left(k \circ \Delta_M^{\circ 3} \circ \partial_{\mu} \sigma\right) V^{'\mu}\right],
\end{equation}
\begin{equation}
\Delta {\cal L}_{mix}^{A'\phi}=\frac{1}{2\varrho^2} \text{tr} \left[2\varrho^2 \left(H^{\left(2\right)} \circ k' \circ \Sigma_M \circ \partial_{\mu} \phi\right) A^{'\mu} + \left(k' \circ \Sigma_M^{\circ 3} \circ \partial_{\mu} \phi\right) A^{'\mu}\right].
\end{equation}
The Hadamard power is used here and it stands for $A^{\circ n} = A \circ A \circ \dots \circ A$, with $A$ appearing $n$ times.

The cancellation of $V\!-\!\sigma$ mixing requires
\begin{align}
& \text{tr}\left\{\left(2\varrho^2 H^{\left(1\right)}\circ k\circ\Delta_M\circ\partial_{\mu}\sigma\right)V^{\mu} - \left(k\circ\Delta_M^{\circ 3}\circ\partial_{\mu}\sigma\right)V^{\mu} - \left(\Delta_M \circ \partial_{\mu} \sigma\right) V^{\mu}\right\} = 0 \nonumber \\
& \Leftrightarrow \ \text{tr} \left\lbrace \left\lbrack \left( 2\varrho^2 H^{\left(1\right)} \circ k \circ \Delta_M  - k \circ \Delta_M^{\circ 3} - \Delta_M \right) \circ \partial_{\mu} \sigma \right\rbrack V^{\mu} \right\rbrace = 0 \nonumber \\
& \Leftrightarrow  \sum_{i,j} \left( 2\varrho^2 H^{\left(1\right)} \circ k \circ \Delta_M  - k \circ \Delta_M^{\circ 3} - \Delta_M \right)_{ij} \partial_{\mu} \sigma_{ij} V^{\mu}_{ji} = 0.
\end{align}
Since the latter sum must vanish, if we equate to zero the coefficients of the independent combination $\partial_{\mu} \sigma_{ij} V^{\mu}_{ji}$ we obtain
\begin{equation}
\left( 2\varrho^2 H^{\left(1\right)} \circ k \circ \Delta_M  - k \circ \Delta_M^{\circ 3} - \Delta_M \right)_{ij} = 0
\end{equation}
for any $i,j$. Due to the antisymmetry of $\Delta_M$, this condition is always satisfied for $i=j$ independently of $k$ values. For $i \neq j$, we have
\begin{eqnarray}
&& 2\varrho^2 H^{\left(1\right)}_{ij} k_{ij} - k_{ij} \left(\Delta_M\right)_{ij}^2 - 1=0 \nonumber \\
&&k_{ij} =  \left(2\varrho^2 H^{\left(1\right)}_{ij} - \left(\Delta_M\right)_{ij}^2\right)^{-1}.
\end{eqnarray}
This expression defines the values of $k$ entries which diagonalize the Lagrangian, and show us that $k$ is a symmetric matrix coinciding (after some renormalizations of fields) with the known result \cite{Ebert86}, for $H^{(1)}\propto \bf 1$. 

A very similar computation may be carried out for $A\!-\!\phi$ mixing, yielding
\begin{equation}
k'_{ij} = \left(2\varrho^2 H^{\left(2\right)}_{ij} + \left(\Sigma_M\right)_{ij}^2\right)^{-1}.
\end{equation}

A convenient way to write these results in matrix form is
\begin{equation}
k = \left(2\varrho^2 H^{\left(1\right)} - \Delta_M^{\circ 2}\right)^{\circ -1}, \qquad
k' = \left(2\varrho^2 H^{\left(2\right)} + \Sigma_M^{\circ 2}\right)^{\circ -1},
\end{equation}
where the Hadamard inverse has been used; its definition may be given as $\left(A^{\circ -1}\right)_{ij} = \left(A_{ij}\right)^{-1}$. It may be checked that these results are in complete agreement with the previously obtained values for $k,k'$ coefficients in \cite{Morais17}. 

We remark that the standard treatment of the problem in \cite{Morais17} requires the analytic manipulation of expressions involving something like 10 or more flavor indices which are contracted among themselves in non-trivial ways. This can easily become a cumbersome and error-prone calculation. Furthermore, the previous form for $X_\mu$ and $Y_\mu$ obscures the fact that each matrix entry of the spin-1 fields is effectively shifted by a single entry of the spin-0 matrix field; this is made explicit within the formalism presented here. The present formulation yields all the results in an efficient and closed-form way.

\section{Conclusion}
Resorting to arguments pertaining to the Lie algebra associated with chiral transformations and to Chisholm's theorem, we have shown that one may always use the most general linear shifts of $V_{\mu}$ and $A_{\mu}$ fields (\ref{shifts}) for dealing with $V\!-\!\sigma$ and $A\!-\!\phi$ mixing in chiral models without compromising the chiral symmetry properties of the Lagrangian, as long as one admits the corresponding new transformation laws (\ref{V-shift-transf},\ref{A-shift-transf}) for the shifted fields. This result is independent of the number of flavors and works even when the $U(n)\times U(n)$ chiral symmetry is explicitly broken to $U(1)^n$. 

Although our arguments have been presented using an NJL-type quark model as a starting point, we expect that our proposed strategy for dealing with the mixing terms is fully applicable to other kinds of chiral models such as the linear sigma model \cite{Gell-Mann60}, massive Yang-Mills models and so on \cite{Meissner88}. This is justified if one understands that the form of these mixing terms is fundamentally constrained by the symmetry requirements which should, in principle, be the same in any effective chiral model for mesons. A particular shifting scheme has been proposed in (\ref{shifts-2}) which is sufficient for dealing with mixing terms appearing in the standard form $\text{tr}\left(V^{\mu} \left\lbrack M,\partial_{\mu}\sigma \right\rbrack\right)$ and $\text{tr}\left(A^{\mu} \left\lbrace M,\partial_{\mu}\phi \right\rbrace\right)$.

\begin{acknowledgments}
	Work supported in part by Funda\c{c}\~{a}o para a Ci\^{e}ncia e a Tecnologia (FCT), through PhD grant SFRH/BD/110315/2015 and project UID/FIS/04564/2016, and Centro de F\'{i}sica da Universidade de Coimbra (CFisUC).
\end{acknowledgments}

\bibliography{mybibfile}

\begin{thebibliography}{41}%
\makeatletter
\providecommand \@ifxundefined [1]{%
 \@ifx{#1\undefined}
}%
\providecommand \@ifnum [1]{%
 \ifnum #1\expandafter \@firstoftwo
 \else \expandafter \@secondoftwo
 \fi
}%
\providecommand \@ifx [1]{%
 \ifx #1\expandafter \@firstoftwo
 \else \expandafter \@secondoftwo
 \fi
}%
\providecommand \natexlab [1]{#1}%
\providecommand \enquote  [1]{``#1''}%
\providecommand \bibnamefont  [1]{#1}%
\providecommand \bibfnamefont [1]{#1}%
\providecommand \citenamefont [1]{#1}%
\providecommand \href@noop [0]{\@secondoftwo}%
\providecommand \href [0]{\begingroup \@sanitize@url \@href}%
\providecommand \@href[1]{\@@startlink{#1}\@@href}%
\providecommand \@@href[1]{\endgroup#1\@@endlink}%
\providecommand \@sanitize@url [0]{\catcode `\\12\catcode `\$12\catcode
  `\&12\catcode `\#12\catcode `\^12\catcode `\_12\catcode `\%12\relax}%
\providecommand \@@startlink[1]{}%
\providecommand \@@endlink[0]{}%
\providecommand \url  [0]{\begingroup\@sanitize@url \@url }%
\providecommand \@url [1]{\endgroup\@href {#1}{\urlprefix }}%
\providecommand \urlprefix  [0]{URL }%
\providecommand \Eprint [0]{\href }%
\providecommand \doibase [0]{http://dx.doi.org/}%
\providecommand \selectlanguage [0]{\@gobble}%
\providecommand \bibinfo  [0]{\@secondoftwo}%
\providecommand \bibfield  [0]{\@secondoftwo}%
\providecommand \translation [1]{[#1]}%
\providecommand \BibitemOpen [0]{}%
\providecommand \bibitemStop [0]{}%
\providecommand \bibitemNoStop [0]{.\EOS\space}%
\providecommand \EOS [0]{\spacefactor3000\relax}%
\providecommand \BibitemShut  [1]{\csname bibitem#1\endcsname}%
\let\auto@bib@innerbib\@empty
\bibitem [{\citenamefont {{S. Coleman}}\ and\ \citenamefont {{E.
  Witten}}(1980)}]{Coleman80}%
  \BibitemOpen
  \bibfield  {author} {\bibinfo {author} {\bibnamefont {{S. Coleman}}}\ and\
  \bibinfo {author} {\bibnamefont {{E. Witten}}},\ }\href {\doibase
  https://doi.org/10.1103/PhysRevLett.45.100} {\bibfield  {journal} {\bibinfo
  {journal} {Phys. Rev. Lett.}\ }\textbf {\bibinfo {volume} {45}},\ \bibinfo
  {pages} {100} (\bibinfo {year} {1980})}\BibitemShut {NoStop}%
\bibitem [{\citenamefont {{Y. Nambu}}(1960)}]{Nambu60}%
  \BibitemOpen
  \bibfield  {author} {\bibinfo {author} {\bibnamefont {{Y. Nambu}}},\ }\href
  {\doibase https://doi.org/10.1103/PhysRevLett.4.380} {\bibfield  {journal}
  {\bibinfo  {journal} {Phys. Rev. Lett.}\ }\textbf {\bibinfo {volume} {4}},\
  \bibinfo {pages} {380} (\bibinfo {year} {1960})}\BibitemShut {NoStop}%
\bibitem [{\citenamefont {{J. Goldstone}}(1961)}]{Goldstone61}%
  \BibitemOpen
  \bibfield  {author} {\bibinfo {author} {\bibnamefont {{J. Goldstone}}},\
  }\href {\doibase 0.1007/BF02812722} {\bibfield  {journal} {\bibinfo
  {journal} {Nouvo Cimento}\ }\textbf {\bibinfo {volume} {19}},\ \bibinfo
  {pages} {154} (\bibinfo {year} {1961})}\BibitemShut {NoStop}%
\bibitem [{\citenamefont {{J. Goldstone}}\ \emph {et~al.}(1962)\citenamefont
  {{J. Goldstone}}, \citenamefont {{A. Salam}},\ and\ \citenamefont {{S.
  Weinberg}}}]{Goldstone62}%
  \BibitemOpen
  \bibfield  {author} {\bibinfo {author} {\bibnamefont {{J. Goldstone}}},
  \bibinfo {author} {\bibnamefont {{A. Salam}}}, \ and\ \bibinfo {author}
  {\bibnamefont {{S. Weinberg}}},\ }\href {\doibase
  https://doi.org/10.1103/PhysRev.127.965} {\bibfield  {journal} {\bibinfo
  {journal} {Phys. Rev.}\ }\textbf {\bibinfo {volume} {127}},\ \bibinfo {pages}
  {965} (\bibinfo {year} {1962})}\BibitemShut {NoStop}%
\bibitem [{\citenamefont {{Y. Nambu}}\ and\ \citenamefont {{G.
  Jona-Lasinio}}(1961{\natexlab{a}})}]{Nambu61a}%
  \BibitemOpen
  \bibfield  {author} {\bibinfo {author} {\bibnamefont {{Y. Nambu}}}\ and\
  \bibinfo {author} {\bibnamefont {{G. Jona-Lasinio}}},\ }\href {\doibase
  10.1103/PhysRev.122.345} {\bibfield  {journal} {\bibinfo  {journal} {Phys.
  Rev.}\ }\textbf {\bibinfo {volume} {122}},\ \bibinfo {pages} {345} (\bibinfo
  {year} {1961}{\natexlab{a}})}\BibitemShut {NoStop}%
\bibitem [{\citenamefont {{Y. Nambu}}\ and\ \citenamefont {{G.
  Jona-Lasinio}}(1961{\natexlab{b}})}]{Nambu61b}%
  \BibitemOpen
  \bibfield  {author} {\bibinfo {author} {\bibnamefont {{Y. Nambu}}}\ and\
  \bibinfo {author} {\bibnamefont {{G. Jona-Lasinio}}},\ }\href {\doibase
  10.1103/PhysRev.124.246} {\bibfield  {journal} {\bibinfo  {journal} {Phys.
  Rev.}\ }\textbf {\bibinfo {volume} {124}},\ \bibinfo {pages} {246} (\bibinfo
  {year} {1961}{\natexlab{b}})}\BibitemShut {NoStop}%
\bibitem [{\citenamefont {{K. Kikkawa}}(1976)}]{Kikkawa76}%
  \BibitemOpen
  \bibfield  {author} {\bibinfo {author} {\bibnamefont {{K. Kikkawa}}},\ }\href
  {\doibase https://doi.org/10.1143/PTP.56.947} {\bibfield  {journal} {\bibinfo
   {journal} {Prog. Theor. Phys.}\ }\textbf {\bibinfo {volume} {56}},\ \bibinfo
  {pages} {947} (\bibinfo {year} {1976})}\BibitemShut {NoStop}%
\bibitem [{\citenamefont {{M. K. Volkov}}(1986)}]{Volkov86}%
  \BibitemOpen
  \bibfield  {author} {\bibinfo {author} {\bibnamefont {{M. K. Volkov}}},\
  }\href {\doibase ISSN 0367-2026} {\bibfield  {journal} {\bibinfo  {journal}
  {PEPAN}\ }\textbf {\bibinfo {volume} {17}},\ \bibinfo {pages} {433} (\bibinfo
  {year} {1986})}\BibitemShut {NoStop}%
\bibitem [{\citenamefont {{D. Ebert}}\ and\ \citenamefont {{H.
  Reinhardt}}(1986)}]{Ebert86}%
  \BibitemOpen
  \bibfield  {author} {\bibinfo {author} {\bibnamefont {{D. Ebert}}}\ and\
  \bibinfo {author} {\bibnamefont {{H. Reinhardt}}},\ }\href {\doibase
  https://doi.org/10.1016/S0550-3213(86)80009-8} {\bibfield  {journal}
  {\bibinfo  {journal} {Nucl. Phys. B}\ }\textbf {\bibinfo {volume} {271}},\
  \bibinfo {pages} {188} (\bibinfo {year} {1986})}\BibitemShut {NoStop}%
\bibitem [{\citenamefont {{J. Bijnens}}\ \emph {et~al.}(1993)\citenamefont {{J.
  Bijnens}}, \citenamefont {{C. Bruno}},\ and\ \citenamefont {{E. de
  Rafael}}}]{Bijnens93}%
  \BibitemOpen
  \bibfield  {author} {\bibinfo {author} {\bibnamefont {{J. Bijnens}}},
  \bibinfo {author} {\bibnamefont {{C. Bruno}}}, \ and\ \bibinfo {author}
  {\bibnamefont {{E. de Rafael}}},\ }\href {\doibase
  https://doi.org/10.1016/0550-3213(93)90466-3} {\bibfield  {journal} {\bibinfo
   {journal} {Nucl. Phys. B}\ }\textbf {\bibinfo {volume} {390}},\ \bibinfo
  {pages} {501} (\bibinfo {year} {1993})}\BibitemShut {NoStop}%
\bibitem [{\citenamefont {{V. Bernard}}\ \emph {et~al.}(1996)\citenamefont {{V.
  Bernard}}, \citenamefont {{A. H. Blin}}, \citenamefont {{B. Hiller}},
  \citenamefont {{Yu. P. Ivanov}}, \citenamefont {{A. A. Osipov}},\ and\
  \citenamefont {{Ulf-G. Meissner}}}]{Bernard96}%
  \BibitemOpen
  \bibfield  {author} {\bibinfo {author} {\bibnamefont {{V. Bernard}}},
  \bibinfo {author} {\bibnamefont {{A. H. Blin}}}, \bibinfo {author}
  {\bibnamefont {{B. Hiller}}}, \bibinfo {author} {\bibnamefont {{Yu. P.
  Ivanov}}}, \bibinfo {author} {\bibnamefont {{A. A. Osipov}}}, \ and\ \bibinfo
  {author} {\bibnamefont {{Ulf-G. Meissner}}},\ }\href {\doibase
  https://doi.org/10.1006/aphy.1996.0081} {\bibfield  {journal} {\bibinfo
  {journal} {Annals of Physics}\ }\textbf {\bibinfo {volume} {249}},\ \bibinfo
  {pages} {499} (\bibinfo {year} {1996})}\BibitemShut {NoStop}%
\bibitem [{\citenamefont {{S. Weinberg}}(1967)}]{Weinberg67}%
  \BibitemOpen
  \bibfield  {author} {\bibinfo {author} {\bibnamefont {{S. Weinberg}}},\
  }\href {\doibase https://doi.org/10.1103/PhysRevLett.18.507} {\bibfield
  {journal} {\bibinfo  {journal} {Phys. Rev. Lett.}\ }\textbf {\bibinfo
  {volume} {18}},\ \bibinfo {pages} {507} (\bibinfo {year} {1967})}\BibitemShut
  {NoStop}%
\bibitem [{\citenamefont {{F. J. Gilman}}\ and\ \citenamefont {{H.
  Harari}}(1968)}]{Gilman68}%
  \BibitemOpen
  \bibfield  {author} {\bibinfo {author} {\bibnamefont {{F. J. Gilman}}}\ and\
  \bibinfo {author} {\bibnamefont {{H. Harari}}},\ }\href {\doibase
  https://doi.org/10.1103/PhysRev.165.1803} {\bibfield  {journal} {\bibinfo
  {journal} {Phys. Rev.}\ }\textbf {\bibinfo {volume} {165}},\ \bibinfo {pages}
  {1803} (\bibinfo {year} {1968})}\BibitemShut {NoStop}%
\bibitem [{\citenamefont {{S. Weinberg}}(1969)}]{Weinberg69}%
  \BibitemOpen
  \bibfield  {author} {\bibinfo {author} {\bibnamefont {{S. Weinberg}}},\
  }\href {\doibase https://doi.org/10.1103/PhysRev.177.2604} {\bibfield
  {journal} {\bibinfo  {journal} {Phys. Rev.}\ }\textbf {\bibinfo {volume}
  {177}},\ \bibinfo {pages} {2604} (\bibinfo {year} {1969})}\BibitemShut
  {NoStop}%
\bibitem [{\citenamefont {{S. Gasiorovicz}}\ and\ \citenamefont {{D. A.
  Geffen}}(1969)}]{Gasiorovicz69}%
  \BibitemOpen
  \bibfield  {author} {\bibinfo {author} {\bibnamefont {{S. Gasiorovicz}}}\
  and\ \bibinfo {author} {\bibnamefont {{D. A. Geffen}}},\ }\href {\doibase
  https://doi.org/10.1103/RevModPhys.41.531} {\bibfield  {journal} {\bibinfo
  {journal} {Rev. of Mod. Phys.}\ }\textbf {\bibinfo {volume} {41}},\ \bibinfo
  {pages} {531} (\bibinfo {year} {1969})}\BibitemShut {NoStop}%
\bibitem [{\citenamefont {{G. Ecker}}\ \emph {et~al.}(1989)\citenamefont {{G.
  Ecker}}, \citenamefont {{J. Gasser}}, \citenamefont {{H. Leutwyler}},
  \citenamefont {{A. Pich}},\ and\ \citenamefont {{E. De Rafael}}}]{Ecker89}%
  \BibitemOpen
  \bibfield  {author} {\bibinfo {author} {\bibnamefont {{G. Ecker}}}, \bibinfo
  {author} {\bibnamefont {{J. Gasser}}}, \bibinfo {author} {\bibnamefont {{H.
  Leutwyler}}}, \bibinfo {author} {\bibnamefont {{A. Pich}}}, \ and\ \bibinfo
  {author} {\bibnamefont {{E. De Rafael}}},\ }\href {\doibase
  https://doi.org/10.1016/0370-2693(89)91627-4} {\bibfield  {journal} {\bibinfo
   {journal} {Phys. Lett. B}\ }\textbf {\bibinfo {volume} {41}},\ \bibinfo
  {pages} {425} (\bibinfo {year} {1989})}\BibitemShut {NoStop}%
\bibitem [{\citenamefont {{M. K. Volkov}}\ and\ \citenamefont {{A. A.
  Osipov}}(2017)}]{Volkov17}%
  \BibitemOpen
  \bibfield  {author} {\bibinfo {author} {\bibnamefont {{M. K. Volkov}}}\ and\
  \bibinfo {author} {\bibnamefont {{A. A. Osipov}}},\ }\href {\doibase
  10.1134/S0021364017040142} {\bibfield  {journal} {\bibinfo  {journal} {JETP
  Letters}\ }\textbf {\bibinfo {volume} {105}},\ \bibinfo {pages} {215}
  (\bibinfo {year} {2017})}\BibitemShut {NoStop}%
\bibitem [{\citenamefont {{A. A. Osipov}}\ and\ \citenamefont {{M. K.
  Volkov}}(2017)}]{Osipov17}%
  \BibitemOpen
  \bibfield  {author} {\bibinfo {author} {\bibnamefont {{A. A. Osipov}}}\ and\
  \bibinfo {author} {\bibnamefont {{M. K. Volkov}}},\ }\href {\doibase
  https://doi.org/10.1016/j.aop.2017.04.011} {\bibfield  {journal} {\bibinfo
  {journal} {Annals of Physics}\ }\textbf {\bibinfo {volume} {382}},\ \bibinfo
  {pages} {50} (\bibinfo {year} {2017})}\BibitemShut {NoStop}%
\bibitem [{\citenamefont {{C. Vafa}}\ and\ \citenamefont {{E.
  Witten}}(1984)}]{Vafa84}%
  \BibitemOpen
  \bibfield  {author} {\bibinfo {author} {\bibnamefont {{C. Vafa}}}\ and\
  \bibinfo {author} {\bibnamefont {{E. Witten}}},\ }\href {\doibase
  https://doi.org/10.1016/0550-3213(84)90230-X} {\bibfield  {journal} {\bibinfo
   {journal} {Nucl. Phys. B}\ }\textbf {\bibinfo {volume} {234}},\ \bibinfo
  {pages} {173} (\bibinfo {year} {1984})}\BibitemShut {NoStop}%
\bibitem [{\citenamefont {{G. 't Hooft}}(1974{\natexlab{a}})}]{Hooft74a}%
  \BibitemOpen
  \bibfield  {author} {\bibinfo {author} {\bibnamefont {{G. 't Hooft}}},\
  }\href {\doibase https://doi.org/10.1016/0550-3213(74)90154-0} {\bibfield
  {journal} {\bibinfo  {journal} {Nucl. Phys. B}\ }\textbf {\bibinfo {volume}
  {72}},\ \bibinfo {pages} {461} (\bibinfo {year}
  {1974}{\natexlab{a}})}\BibitemShut {NoStop}%
\bibitem [{\citenamefont {{G. 't Hooft}}(1974{\natexlab{b}})}]{Hooft74b}%
  \BibitemOpen
  \bibfield  {author} {\bibinfo {author} {\bibnamefont {{G. 't Hooft}}},\
  }\href {\doibase https://doi.org/10.1016/0550-3213(74)90088-1} {\bibfield
  {journal} {\bibinfo  {journal} {Nucl. Phys. B}\ }\textbf {\bibinfo {volume}
  {75}},\ \bibinfo {pages} {461} (\bibinfo {year}
  {1974}{\natexlab{b}})}\BibitemShut {NoStop}%
\bibitem [{\citenamefont {{E. Witten}}(1979)}]{Witten79}%
  \BibitemOpen
  \bibfield  {author} {\bibinfo {author} {\bibnamefont {{E. Witten}}},\ }\href
  {\doibase https://doi.org/10.1016/0550-3213(79)90232-3} {\bibfield  {journal}
  {\bibinfo  {journal} {Nucl. Phys. B}\ }\textbf {\bibinfo {volume} {160}},\
  \bibinfo {pages} {57} (\bibinfo {year} {1979})}\BibitemShut {NoStop}%
\bibitem [{\citenamefont {{Ulf-G. Meissner}}(1988)}]{Meissner88}%
  \BibitemOpen
  \bibfield  {author} {\bibinfo {author} {\bibnamefont {{Ulf-G. Meissner}}},\
  }\href {\doibase https://doi.org/10.1016/0370-1573(88)90090-7} {\bibfield
  {journal} {\bibinfo  {journal} {Physics Reports}\ }\textbf {\bibinfo {volume}
  {161}},\ \bibinfo {pages} {213} (\bibinfo {year} {1988})}\BibitemShut
  {NoStop}%
\bibitem [{\citenamefont {{M. C. Birse}}(1996)}]{Birse96}%
  \BibitemOpen
  \bibfield  {author} {\bibinfo {author} {\bibnamefont {{M. C. Birse}}},\
  }\href {\doibase 0.1007/s002180050105} {\bibfield  {journal} {\bibinfo
  {journal} {Zeitschrift f\" ur Physik A Hadrons and Nuclei}\ }\textbf
  {\bibinfo {volume} {355}},\ \bibinfo {pages} {231} (\bibinfo {year}
  {1996})}\BibitemShut {NoStop}%
\bibitem [{\citenamefont {{A. A. Osipov}}\ and\ \citenamefont {{B.
  Hiller}}(2000)}]{Osipov00}%
  \BibitemOpen
  \bibfield  {author} {\bibinfo {author} {\bibnamefont {{A. A. Osipov}}}\ and\
  \bibinfo {author} {\bibnamefont {{B. Hiller}}},\ }\href {\doibase
  https://doi.org/10.1103/PhysRevD.62.114013} {\bibfield  {journal} {\bibinfo
  {journal} {Phys. Rev. D}\ }\textbf {\bibinfo {volume} {62}},\ \bibinfo
  {pages} {114013} (\bibinfo {year} {2000})}\BibitemShut {NoStop}%
\bibitem [{\citenamefont {{T. Hatsuda}}\ and\ \citenamefont {{T.
  Kunihiro}}(1994)}]{Hatsuda94}%
  \BibitemOpen
  \bibfield  {author} {\bibinfo {author} {\bibnamefont {{T. Hatsuda}}}\ and\
  \bibinfo {author} {\bibnamefont {{T. Kunihiro}}},\ }\href {\doibase
  https://doi.org/10.1016/0370-1573(94)90022-1} {\bibfield  {journal} {\bibinfo
   {journal} {Physics Reports}\ }\textbf {\bibinfo {volume} {247}},\ \bibinfo
  {pages} {221} (\bibinfo {year} {1994})}\BibitemShut {NoStop}%
\bibitem [{\citenamefont {{Masako Bando}}\ \emph {et~al.}(1988)\citenamefont
  {{Masako Bando}}, \citenamefont {{Taichiro Kugo}},\ and\ \citenamefont
  {{Koichi Yamawaki}}}]{Bando88}%
  \BibitemOpen
  \bibfield  {author} {\bibinfo {author} {\bibnamefont {{Masako Bando}}},
  \bibinfo {author} {\bibnamefont {{Taichiro Kugo}}}, \ and\ \bibinfo {author}
  {\bibnamefont {{Koichi Yamawaki}}},\ }\href {\doibase
  https://doi.org/10.1016/0370-1573(88)90019-1} {\bibfield  {journal} {\bibinfo
   {journal} {Physics Reports}\ }\textbf {\bibinfo {volume} {164}},\ \bibinfo
  {pages} {217} (\bibinfo {year} {1988})}\BibitemShut {NoStop}%
\bibitem [{\citenamefont {Volkov}\ \emph {et~al.}(2012)\citenamefont {Volkov},
  \citenamefont {Arbuzov},\ and\ \citenamefont {Kostunin}}]{Volkov2012tau}%
  \BibitemOpen
  \bibfield  {author} {\bibinfo {author} {\bibfnamefont {M.~K.}\ \bibnamefont
  {Volkov}}, \bibinfo {author} {\bibfnamefont {A.~B.}\ \bibnamefont {Arbuzov}},
  \ and\ \bibinfo {author} {\bibfnamefont {D.~G.}\ \bibnamefont {Kostunin}},\
  }\href {\doibase 10.1103/PhysRevD.86.057301} {\bibfield  {journal} {\bibinfo
  {journal} {Phys. Rev. D}\ }\textbf {\bibinfo {volume} {86}},\ \bibinfo
  {pages} {057301} (\bibinfo {year} {2012})}\BibitemShut {NoStop}%
\bibitem [{\citenamefont {Volkov}\ and\ \citenamefont
  {Kostunin}(2012)}]{Volkov2012elpos}%
  \BibitemOpen
  \bibfield  {author} {\bibinfo {author} {\bibfnamefont {M.~K.}\ \bibnamefont
  {Volkov}}\ and\ \bibinfo {author} {\bibfnamefont {D.~G.}\ \bibnamefont
  {Kostunin}},\ }\href {\doibase 10.1103/PhysRevC.86.025202} {\bibfield
  {journal} {\bibinfo  {journal} {Phys. Rev. C}\ }\textbf {\bibinfo {volume}
  {86}},\ \bibinfo {pages} {025202} (\bibinfo {year} {2012})}\BibitemShut
  {NoStop}%
\bibitem [{\citenamefont {Beni{\'{c}}}(2014)}]{Benic2014}%
  \BibitemOpen
  \bibfield  {author} {\bibinfo {author} {\bibfnamefont {S.}~\bibnamefont
  {Beni{\'{c}}}},\ }\href {\doibase 10.1140/epja/i2014-14111-1} {\bibfield
  {journal} {\bibinfo  {journal} {The European Physical Journal A}\ }\textbf
  {\bibinfo {volume} {50}},\ \bibinfo {pages} {111} (\bibinfo {year}
  {2014})}\BibitemShut {NoStop}%
\bibitem [{\citenamefont {Chu}\ \emph {et~al.}(2015)\citenamefont {Chu},
  \citenamefont {Wang}, \citenamefont {Chen},\ and\ \citenamefont
  {Huang}}]{Chu2015}%
  \BibitemOpen
  \bibfield  {author} {\bibinfo {author} {\bibfnamefont {P.-C.}\ \bibnamefont
  {Chu}}, \bibinfo {author} {\bibfnamefont {X.}~\bibnamefont {Wang}}, \bibinfo
  {author} {\bibfnamefont {L.-W.}\ \bibnamefont {Chen}}, \ and\ \bibinfo
  {author} {\bibfnamefont {M.}~\bibnamefont {Huang}},\ }\href {\doibase
  10.1103/PhysRevD.91.023003} {\bibfield  {journal} {\bibinfo  {journal} {Phys.
  Rev. D}\ }\textbf {\bibinfo {volume} {91}},\ \bibinfo {pages} {023003}
  (\bibinfo {year} {2015})}\BibitemShut {NoStop}%
\bibitem [{\citenamefont {Gatto}\ and\ \citenamefont
  {Ruggieri}(2011)}]{Gatto2010}%
  \BibitemOpen
  \bibfield  {author} {\bibinfo {author} {\bibfnamefont {R.}~\bibnamefont
  {Gatto}}\ and\ \bibinfo {author} {\bibfnamefont {M.}~\bibnamefont
  {Ruggieri}},\ }\href {\doibase 10.1103/PhysRevD.83.034016} {\bibfield
  {journal} {\bibinfo  {journal} {Phys. Rev. D}\ }\textbf {\bibinfo {volume}
  {83}},\ \bibinfo {pages} {034016} (\bibinfo {year} {2011})}\BibitemShut
  {NoStop}%
\bibitem [{\citenamefont {Moreira}\ \emph {et~al.}(2015)\citenamefont
  {Moreira}, \citenamefont {Morais}, \citenamefont {Hiller}, \citenamefont
  {Osipov},\ and\ \citenamefont {Blin}}]{Moreira2014}%
  \BibitemOpen
  \bibfield  {author} {\bibinfo {author} {\bibfnamefont {J.}~\bibnamefont
  {Moreira}}, \bibinfo {author} {\bibfnamefont {J.}~\bibnamefont {Morais}},
  \bibinfo {author} {\bibfnamefont {B.}~\bibnamefont {Hiller}}, \bibinfo
  {author} {\bibfnamefont {A.~A.}\ \bibnamefont {Osipov}}, \ and\ \bibinfo
  {author} {\bibfnamefont {A.~H.}\ \bibnamefont {Blin}},\ }\href {\doibase
  10.1103/PhysRevD.91.116003} {\bibfield  {journal} {\bibinfo  {journal} {Phys.
  Rev. D}\ }\textbf {\bibinfo {volume} {91}},\ \bibinfo {pages} {116003}
  (\bibinfo {year} {2015})}\BibitemShut {NoStop}%
\bibitem [{\citenamefont {Pais}\ \emph {et~al.}(2016)\citenamefont {Pais},
  \citenamefont {Menezes},\ and\ \citenamefont {Provid\^encia}}]{Pais2016}%
  \BibitemOpen
  \bibfield  {author} {\bibinfo {author} {\bibfnamefont {H.}~\bibnamefont
  {Pais}}, \bibinfo {author} {\bibfnamefont {D.~P.}\ \bibnamefont {Menezes}}, \
  and\ \bibinfo {author} {\bibfnamefont {C.}~\bibnamefont {Provid\^encia}},\
  }\href {\doibase 10.1103/PhysRevC.93.065805} {\bibfield  {journal} {\bibinfo
  {journal} {Phys. Rev. C}\ }\textbf {\bibinfo {volume} {93}},\ \bibinfo
  {pages} {065805} (\bibinfo {year} {2016})}\BibitemShut {NoStop}%
\bibitem [{\citenamefont {{R. D. Ball}}(1989)}]{Ball89}%
  \BibitemOpen
  \bibfield  {author} {\bibinfo {author} {\bibnamefont {{R. D. Ball}}},\ }\href
  {\doibase https://doi.org/10.1016/0370-1573(89)90027-6} {\bibfield  {journal}
  {\bibinfo  {journal} {Physics Reports}\ }\textbf {\bibinfo {volume} {182}},\
  \bibinfo {pages} {1} (\bibinfo {year} {1989})}\BibitemShut {NoStop}%
\bibitem [{\citenamefont {{I. S. R. Chisholm}}(1961)}]{Chisholm61}%
  \BibitemOpen
  \bibfield  {author} {\bibinfo {author} {\bibnamefont {{I. S. R. Chisholm}}},\
  }\href {\doibase https://doi.org/10.1016/0029-5582(61)90106-7} {\bibfield
  {journal} {\bibinfo  {journal} {Nucl. Phys.}\ }\textbf {\bibinfo {volume}
  {26}},\ \bibinfo {pages} {469} (\bibinfo {year} {1961})}\BibitemShut
  {NoStop}%
\bibitem [{\citenamefont {{S. Kamefuchi}}\ \emph {et~al.}(1961)\citenamefont
  {{S. Kamefuchi}}, \citenamefont {{L. O'Raifeartaigh}},\ and\ \citenamefont
  {{Abdus Salam}}}]{Salam61}%
  \BibitemOpen
  \bibfield  {author} {\bibinfo {author} {\bibnamefont {{S. Kamefuchi}}},
  \bibinfo {author} {\bibnamefont {{L. O'Raifeartaigh}}}, \ and\ \bibinfo
  {author} {\bibnamefont {{Abdus Salam}}},\ }\href {\doibase
  https://doi.org/10.1016/0029-5582(61)90056-6} {\bibfield  {journal} {\bibinfo
   {journal} {Nucl. Phys.}\ }\textbf {\bibinfo {volume} {28}},\ \bibinfo
  {pages} {529} (\bibinfo {year} {1961})}\BibitemShut {NoStop}%
\bibitem [{\citenamefont {Styan}(1973)}]{Styan73}%
  \BibitemOpen
  \bibfield  {author} {\bibinfo {author} {\bibfnamefont {G.~P.}\ \bibnamefont
  {Styan}},\ }\href {\doibase http://dx.doi.org/10.1016/0024-3795(73)90023-2}
  {\bibfield  {journal} {\bibinfo  {journal} {Linear Algebra and its
  Applications}\ }\textbf {\bibinfo {volume} {6}},\ \bibinfo {pages} {217}
  (\bibinfo {year} {1973})}\BibitemShut {NoStop}%
\bibitem [{\citenamefont {{J. Morais}}\ \emph {et~al.}(2017)\citenamefont {{J.
  Morais}}, \citenamefont {{B. Hiller}},\ and\ \citenamefont {{A. A.
  Osipov}}}]{Morais17}%
  \BibitemOpen
  \bibfield  {author} {\bibinfo {author} {\bibnamefont {{J. Morais}}}, \bibinfo
  {author} {\bibnamefont {{B. Hiller}}}, \ and\ \bibinfo {author} {\bibnamefont
  {{A. A. Osipov}}},\ }\href {\doibase
  https://doi.org/10.1103/PhysRevD.95.074033} {\bibfield  {journal} {\bibinfo
  {journal} {Physical Review D}\ }\textbf {\bibinfo {volume} {95}},\ \bibinfo
  {pages} {074033} (\bibinfo {year} {2017})}\BibitemShut {NoStop}%
\bibitem [{\citenamefont {{G. 't Hooft}}(1976)}]{tHooft76}%
  \BibitemOpen
  \bibfield  {author} {\bibinfo {author} {\bibnamefont {{G. 't Hooft}}},\
  }\href {\doibase 10.1103/PhysRevLett.37.8} {\bibfield  {journal} {\bibinfo
  {journal} {Phys. Rev. Lett.}\ }\textbf {\bibinfo {volume} {37}},\ \bibinfo
  {pages} {8} (\bibinfo {year} {1976})}\BibitemShut {NoStop}%
\bibitem [{\citenamefont {{M. Gell-Mann}}\ and\ \citenamefont {{M.
  L{\'e}vy}}(1960)}]{Gell-Mann60}%
  \BibitemOpen
  \bibfield  {author} {\bibinfo {author} {\bibnamefont {{M. Gell-Mann}}}\ and\
  \bibinfo {author} {\bibnamefont {{M. L{\'e}vy}}},\ }\href {\doibase
  10.1007/BF02859738} {\bibfield  {journal} {\bibinfo  {journal} {Il Nuovo
  Cimento (1955-1965)}\ }\textbf {\bibinfo {volume} {16}},\ \bibinfo {pages}
  {705} (\bibinfo {year} {1960})}\BibitemShut {NoStop}%
\end{thebibliography}%
\end{document}